\documentclass[twocolumn,showpacs,amsmath,amssymb,amsfonts,nofootinbib,plb]{revtex4-1}
\usepackage{graphicx}% Include figure files
\usepackage{dcolumn}% Align table columns on decimal point
\usepackage{bm}% bold math
\usepackage{ulem}% bold math
\usepackage{overpic}

\newcommand{\nn}{\nonumber}

\def\beq{\begin{equation}}
\def\eeq{\end{equation}}

\usepackage{soul}
\usepackage{cancel}

\usepackage[usenames]{color}

\begin{document}
\input{epsf}

\title{Small-$x$ evolution of jet quenching parameter}

 \author{Raktim Abir}
 %\affiliation{Theory Division, Saha Institute of Nuclear Physics,  
 %1/AF Bidhannagar, Kolkata 700064, India.}
 \affiliation{Department of Physics and Astronomy, Wayne State University, 
 666 W. Hancock St., Detroit, MI 48201, USA.}

\begin{abstract}
 Concept of transverse deflection probability of a parton that travels through strongly interacting medium, 
 recently introduced by D'Eramo, Liu and Rajagopal, have been used to derive high energy evolution equation
 for the jet quenching parameter in stochastic multiple scatterings regime. 
 %which is the region of interest
 %in the context of jet quenching phenomenology of the heavy-ion collider experiments. 
 Jet quenching parameter, $\hat q(x)$, appears to evolve with $x$, with an exponent $0.9{\bar \alpha}_s$, which is slightly less than 
 that of $x {\cal G} (x)$ where ${\cal G} (x)$ is the gluon distribution function.   
\end{abstract}


\date{\today}
\maketitle

 %%%%%%%%%%%%%%%%%%%%%%%%%%%%%%%%%%%%%%%%%%%%%%%%%%%%%%%%%%%%%%%%%%%%%%%%%%%%%%%%%%%%%%%%%%%%%%%%%%%%
 \section{Introduction}
 %%%%%%%%%%%%%%%%%%%%%%%%%%%%%%%%%%%%%%%%%%%%%%%%%%%%%%%%%%%%%%%%%%%%%%%%%%%%%%%%%%%%%%%%%%%%%%%%%%%%
 %
 The heavy-ion collision programs at CERN's Large Hadron Collider have opened new access for 
 exploration of extreme hot and dense nuclear matter.   
 Precision tomography of the nuclear matter are now become feasible as accurate test of the 
 underlying quantum chromodynamic (QCD) theory.  
 This is instrumental in discovering yet unexplored characteristics of various nuclear effects and 
 collective phenomena that the nuclear matter may possess. 
 One possibility is to explore the QCD scale/energy evolution of various observables in this 
 extreme ambiance. 
 Advancement of study for hard sector observables at the LHC elevated the medium modification of 
 high energy jets as prevailing topic of investigation.  
 In this context study on scale/energy evolution of the jet quenching parameter, attributed as 
 the stopping power of the medium for certain probe of the medium, is now viable. 
 %
 %
 %%%%%%%%%%%%%%%%%%%%%%%%%%%%%%%%%%%%%%%%%%%%%%%%%%%%%%%%%%%%%%%%%%%%%%%%%%%%%%%%%%%%%%%%%%%%%%%%%%%%
 %
 High energy quarks and gluons passing through the interacting nuclear matter have their transverse 
 momentum distribution broadened due to multiple scatterings with the constituents of the medium.  
 While travelling through the strongly coupled medium the hard parton loses energy as well as its
 direction of momentum changes. 
 Change in the direction of momentum is referred as `transverse momentum broadening' for the 
 travelling parton.
 In the context of jet-medium interaction the broadening refers to the effect on the jet when the 
 direction of the momenta of an ensemble of partons changes due to the random kicks. 
 Even though there is no apparent change in mean momenta, the spread of the momentum distribution
 of individual parton within that ensemble broaden. 
 %
 
 %%%%%%%%%%%%%%%%%%%%%%%%%%%%%%%%%%%%%%%%%%%%%%%%%%%%%%%%%%%%%%%%%%%%%%%%%%%%%%%%%%%%%%%%%%%%%%%%%%%%
 %
 Evolution of the momentum broadening was first studied by Liou, Mueller and Wu \cite{Liou:2013qya} 
 by introducing radiative modification over the leading order momentum broadening effect.
 The authors shown that average momentum broadening $\langle p_{\perp}^{2} \rangle$ has both double 
 and single logarithmic terms. 
 Both the double logarithmic terms, $\ln^2(L/l_0)$, and single logarithmic terms, $\ln(L/l_0)$ are
 coming from gluon radiation induced by the medium interactions. 
 Here length of the nuclear matter is $L$ and $l_0$ being the size of constituents of the matter.
 Their estimation shown that the radiative contribution to be a sizeable correction to the 
 nonradiative leading value of $\langle p_{\perp}^{2} \rangle$. 
 Later an evolution equation have been obtained for the inclusive one-gluon distribution, through 
 the concept of classical branching process and cascade of partons \cite{Blaizot:2013vha}. 
 This explicitly takes into account the dependence of the observed gluon spectrum upon the energy 
 and the transverse momentum.
 The explicit transverse momentum dependence of the splitting kernel then enable one to identify large 
 corrections to the jet quenching parameter. 
 Subsequent studies on non-linear evolution leads to prescribe the renormalization of the jet-quenching 
 parameter \cite{Iancu:2014kga,Blaizot:2014bha}.  
 %
 %%%%%%%%%%%%%%%%%%%%%%%%%%%%%%%%%%%%%%%%%%%%%%%%%%%%%%%%%%%%%%%%%%%%%%%%%%%%%%%%%%%%%%%%%%%%%%%%%%%%
  
 In this paper, in order to study the energy evolution of jet quenching parameter, we have adopted 
 the idea of transverse deflection probability of a parton, that travels through the nuclear medium.  
 Following a recent work by D'Eramo, Liu and Rajagopal \cite{D'Eramo:2010ak} we then relate the 
 momentum broadening to the $S$-matrx of the nuclear interaction for a dipole. 
 Balitsky-Kovchegov equation as the evolution equation of the $S$-matrix is then used to derive high 
 energy evolution equation for the jet quenching parameter in stochastic multiple scatterings regime. 
 The known result of double log enhancement emerges as a special case in the limit when the single
 scattering is only contributing. 
 Power-counting techniques borrowed from Soft-Collinear-Effective-Theory (SCET)~
 \cite{Bauer:2000yr,Bauer:2001ct,Bauer:2002nz,Bauer:2001yt} have been used to identify the leading contributions in 
 the stochastic multiple scatterings region. 
 For an almost constant ${\hat q}(\omega)$ we recovered the double log result (in the limit 
 $Q_s^2 \rightarrow {\hat q}L$) first derived in \cite{Liou:2013qya} and subsequent other studies 
 \cite{Blaizot:2013vha,Iancu:2014kga,Blaizot:2014bha}.   
 We have also shown that the double log enhancement diluted when we go beyond single scattering limit 
 as previously argued in \cite{Liou:2013qya}. 
 Jet quenching parameter $\hat q(x)$ is found to evolve with $x$, slightly weaker than $x {\cal G} (x)$ 
 where ${\cal G} (x)$ is the gluon distribution function.  
 
 \section{Probability density for transverse deflection $P(k_\perp)$}
 The transverse momentum broadening of a parton can be studied by introducing concept of a probability 
 density, denoted in this article as $P(k_\perp)$. 
 It signifies the probabilistic weight for the event where, after travelling a medium of 
 length $L$, amount of transverse momentum that the parton acquired is $k_\perp$ \cite{D'Eramo:2010ak}. 
 The probability density $P(k_\perp)$ is chosen to be normalized in the following way, 
 \begin{eqnarray}
 \int \frac{d^2k_\perp}{(2\pi)^2}P(k_\perp)=
 \frac{1}{4\pi} \int dk_\perp^2~P(k_\perp)=1 ~.
 \label{prob}
 \end{eqnarray}
 \noindent Using this probability density one can quickly estimate mean transverse 
 momentum picked up by the hard parton per unit distance travelled, 
 \begin{eqnarray}
 {\hat q} \equiv \frac{\langle k_\perp^2 \rangle}{L} = \frac{1}{4\pi L} \int dk_\perp^2~k_\perp^2~P(k_\perp) ~.
 \label{sayarabanu}
 \end{eqnarray}
 This defines the jet broadening (or quenching) parameter ${\hat q}$
 that may have intrinsic dependence on $x$ and $Q$ of the probe through $P(k_\perp)$. 

 Its important to make certain caveat when defining the jet quenching parameter, ${\hat q}$, 
 as done in  Eq. (\ref{sayarabanu}). In absence of any finite upper bound to the integral, the right-hand side 
 of Eq. (\ref{sayarabanu}) could be diverging, e.g, when $P(k_\perp)$ behaves as power law, 
 $P(k_\perp) \sim 1/k_\perp^4$ in the event of lowest order perturbative gluon production at large-$k_\perp$. 
 This gives a logarithmic UV divergence, resulting both  average transverse momentum $\langle k_\perp^2 \rangle$ as well as 
 jet quenching parameter, ${\hat q}$, infinite. Although, in case of multiple scattering with Sudakov like form factor 
 or in the event of multiple stochastic scatterings where $P(k_\perp)$ takes the form of an exponentially damping factor, 
 right hand side of Eq. (\ref{sayarabanu}) should be finite even without any apparent  upper bound in the integral. 

 In this paper we will discuss the event of multiple stochastic scatterings, where the probe receives random transverse 
 kicks and $P(k_\perp)$ takes the form of a Gaussian with
 a variance of ${\hat q}L/2$,  
 \begin{eqnarray}
 P(k_\perp)= \frac{4\pi}{{\hat q}L}~\exp\left(-\frac{k_\perp^2}{{\hat q}L}\right)~.
 \label{divya}
 \end{eqnarray}
 
 %%%%%%%%%%%%%%%%%%%% %%%%%%%%%%%%%%%%%%%% %%%%%%%%%%%%%%%%%%%% %%%%%%%%%%%%%%%%%%%% %%%%%%%%%%%%%%%%%%%%
 %
 In this study we are looking for transverse momentum broadening of the hard parton that have initial 
 light cone momentum $p(p^+,p^-,p_\perp) \sim Q(0,1,0)$ and enters in a brick of strongly interacting medium of length $L$.
 In order to do the power counting, we have introduced the dimensionless small
 parameter $\lambda$. 
 %
 % \checkmark %%%%%%%%%%%%%%%%%%%%%%%%%%%%%%%%%%%%%%%%%%%%%%%%%%%%%%%%%%%%%%%%%%%%%%%%%%%%%%%%%%%%%%%
 Power corrections in $\lambda$ to some hard process are generally suppressed in the presence of a hard scale,
 $Q^2 \gg \Lambda_{QCD}^2$, which act as base for the power corrections.
 %
 % \checkmark %%%%%%%%%%%%%%%%%%%%%%%%%%%%%%%%%%%%%%%%%%%%%%%%%%%%%%%%%%%%%%%%%%%%%%%%%%%%%%%%%%%%%%%
 This is a concept borrowed from soft collinear effective theory (SCET) studies \cite{Bauer:2000yr,
 Bauer:2001ct,Bauer:2002nz,Bauer:2001yt}.
 %
 % \checkmark %%%%%%%%%%%%%%%%%%%%%%%%%%%%%%%%%%%%%%%%%%%%%%%%%%%%%%%%%%%%%%%%%%%%%%%%%%%%%%%%%%%%%%%
 In SCET studies the power counting protocol is as follows, 
 %
 % \checkmark %%%%%%%%%%%%%%%%%%%%%%%%%%%%%%%%%%%%%%%%%%%%%%%%%%%%%%%%%%%%%%%%%%%%%%%%%%%%%%%%%%%%%%%
 terms that are sub-leading in two order of magnitude can be dropped but the terms that are 
 suppressed by one order of magnitude should be kept. 
 %
 %
 % \checkmark %%%%%%%%%%%%%%%%%%%%%%%%%%%%%%%%%%%%%%%%%%%%%%%%%%%%%%%%%%%%%%%%%%%%%%%%%%%%%%%%%%%%%%%
 %The scaling variable $\lambda$ in chosen such a way that perturbation theory may be applied 
 %down to momentum transfer scales at or above $\lambda^{3/2}Q\sim \Lambda_{\rm QCD}$.
 %
 While travelling through nuclear medium the hard partons interacts repeatedly with the Glauber gluons which scale as 
 $Q(\lambda^2,\lambda^2,\lambda)$. 
 After first a few scatterings the hard parton's momentum becomes of order $Q(\lambda^2,1,\lambda)$, 
 though still quite collinear.
 In this scenario of high energy regime where Glauber gluons are mostly effective, it has been   
 shown in Ref. \cite{D'Eramo:2010ak} that $P(k_\perp)$ can be expressed as the Fourier 
 transform in $r_\perp$ of the expectation value of two light-like path-ordered Wilson lines 
 transversely separated by $r_\perp$,
 \begin{eqnarray}
 P(k_\perp) = \int d^2r_\perp ~ e^{-ik_\perp r_\perp}~{\cal S}(r_\perp)~,
 \label{kate_winslet}
 \end{eqnarray}
 where, 
 \begin{eqnarray}
 {\cal S}(r_\perp) \equiv \frac{1}{N_c} \langle ~{\rm Tr} \left[{\cal W} \left(0,r_\perp\right) 
 {\cal W} \left(0,0\right)\right]~ \rangle ~,
 \end{eqnarray}
 \noindent with, 
 \begin{eqnarray}
 {\cal W}(y^+,y_\perp) \equiv {\cal P} \left\{ \exp\left[ig\int_0^{L^-} dy^- A^+\left(y^+,y^-,y_\perp\right)\right] \right\}~.
 \label{edmond} \nn \\
 \end{eqnarray}
 Within a dipole picture  transverse separation $y_\perp - y_\perp'$  can be taken as the transverse size 
 of the dipole $r_\perp$. The length of the medium is $L = L^{-}/{\sqrt 2}$. 
 In this work we have assumed that $P(k_\perp)$ as the 
 Fourier transform of ${\cal S}(r_\perp)$ \label{kate_winslet} is well valid   
 while in high energy regime $i.e.$ at small-$x$ and 
 all the evolution characteristics for $P(k_\perp)$ exclusively contained in ${\cal S}(r_\perp)$,
 so that we may write, 
 \begin{eqnarray}
 P(k_\perp, Y) = \int d^2r_\perp ~ e^{-ik_\perp r_\perp}~{\cal S}(r_\perp, Y)~,
 \label{kate_winslet20}  
 \end{eqnarray}
 and,
 \begin{eqnarray}
 \frac{\partial P(k_\perp, Y)}{\partial Y} = \int d^2r_\perp ~ e^{-ik_\perp r_\perp}~\frac{\partial {\cal S}(k_\perp, Y)}{\partial Y}~,
 \label{kate_winslet2}
 \end{eqnarray}
 and therefore from Eq. \eqref{sayarabanu}, 
 \begin{eqnarray}
 \frac{\partial {\hat q}(Y)}{\partial Y}&=& \frac{1}{4\pi L} \int dk_\perp^2~k_\perp^2~ \int d^2r_\perp 
           ~ e^{-ik_\perp r_\perp}~  \frac{\partial {\cal S}(r_\perp,Y)}{\partial Y}  \nn \\
  &=& {\hat {\cal F}}_{[k_\perp, r_\perp]}
            \left[ \frac{\partial {\cal S}(r_\perp,Y)}{\partial Y}\right]~ \label{helen}. 
 \end{eqnarray}
 
 \noindent Where for brevity in notations we have introduced ${\hat {\cal F}}$, 
 \begin{eqnarray}
 {\hat {\cal F}}_{[k_\perp, r_\perp]}[{\cal O}] \equiv \frac{1}{4\pi L} \int dk_\perp^2~k_\perp^2~ \int d^2r_\perp 
           ~ e^{-ik_\perp r_\perp}~ {\cal O}~.
  \label{Farida}
 \end{eqnarray}
 It can also be shown that, 
 \begin{eqnarray}
  {\hat {\cal F}}_{[k_\perp, r_\perp]}\left[ r_\perp^2 
                       \exp\left(-\frac{{\hat q}}{4}Lr_\perp^2\right)\right] &=& -\frac{4}{L}~,  \label{waheeda1} \\
  {\hat {\cal F}}_{[k_\perp, r_\perp]}\left[ r_\perp^4 
                       \exp\left(-\frac{{\hat q}}{4}Lr_\perp^2\right)\right] &=& 0~.  \label{waheeda2}  
  \label{waheeda1}
 \end{eqnarray}
 Now, the non-linear  evolution of the $S$-matrix, ${\cal S}(r_\perp=y_\perp-y_\perp')$, is governed  
 by the Balitsky-Kovchegov equation (BK)   
 \cite{Balitsky:1995ub,Kovchegov:1999ua,Kovchegov:1999yj,Gelis:2012ri}, in large-$N_c$ limit as,   
 \begin{eqnarray}
 \frac{\partial {\cal S}(y_\perp,y_\perp';Y)}{\partial Y} &=& - \frac{\alpha_s N_c}{2\pi^2} 
               \int d^2z_\perp \frac{(y_\perp-y'_\perp)^2}{(y_\perp-z_\perp)^2(z_\perp-y_\perp')^2}  \nn \\
  &&             \left[{\cal S}(y_\perp,y_\perp';Y)-{\cal S}(y_\perp,z_\perp;Y){\cal S}(z_\perp,y_\perp',Y) \right]~. \nn \\
 \label{suchitra}
 \end{eqnarray}
 Using Eq. \eqref{helen} and Eq. \eqref{suchitra} with $r_\perp=y_\perp-y_\perp'$, together with  
 fact that the medium is transnationally invariant, evolution equation for ${\hat q}(Y)$, can now
 be written as, 
 \begin{eqnarray}
 \frac{\partial {\hat q(Y)}}{\partial Y} 
 &=& -\frac{\alpha_s N_c}{2\pi^2} {\hat{\cal F}}_{[k_\perp, r_\perp]}[{\cal M}(r_\perp)] ~,
 \label{waheeda100}
 \end{eqnarray}
 where ${\cal M}(r_\perp)$  is an integral over the daughter dipoles' transverse coordinates, 
 \begin{eqnarray}
  {\cal M}(r_\perp) &=& \int d^2B_\perp    \frac{r_\perp^2}{(r_\perp-B_\perp)^2B_\perp^2}  \nn \\
  && \left[{\cal S}(r_\perp, Y) -{\cal S}(r_\perp-B_\perp, Y) {\cal S}(B_\perp, Y) \right]
  \label{kajal}
 \end{eqnarray}
 The Balitsky-Kovchegov (BK) equation results from summing up long-lived soft gluon emissions,
 off almost onshell hard quark, where the gluon lifetimes usually much longer than the size 
 of the target, which is taken to be a brick of size $L$ here. 
 The integral in Eq. \eqref{edmond} should therefore run from $-\infty$ to $+\infty$, which means that the 
 quark, scattering on the brick, is created at time, $\tau_i =-\infty$, flies through semi-infinite empty 
 space almost onshell while developing the gluon cascade giving the evolution of Eq. \eqref{suchitra}, 
 scatters on the brick, and flies off until $\tau_f =+\infty$ again developing the cascade \cite{Dumitru:2002qt}. 
 Such a physical picture certainly does not apply in the context of jet quenching in heavy ion collisions
 where quark jets, are produced in the actual collision (at $\tau=0$), with a high virtuality $Q$ and quickly losses 
 the virtuality by ordered emission through (medium modified) DGLAP \cite{Dokshitzer:1977sg,Gribov:1972ri,Altarelli:1977zs}
 evolution and have no time to develop a 
 gluon cascade accordance with Eq. \eqref{suchitra} before interacting with the plasma.
 In the present discussion we therefore presume ${\hat q}$ as momentum broadening long after the quark 
 leaves the brick when it again becomes almost onshell. 

 In the event when $Y \sim 0$ the jet quenching parameter ${\hat q}_{0}$ and the saturation momentum  $Q_{s0}$ are 
 approximately related as ${\hat q}_{0}\sim 4 Q_{s0}^2/L$.
 In the rest of the article we will evaluate Eq. \eqref{waheeda100} inside or around saturation region 
 which is the region of interest relevant in the context of transverse momentum broadening and 
 energy loss phenomenology in relativistic heavy-ion collision studies.   
 
 %%%%%%%%%%%%%%%%%%%% %%%%%%%%%%%%%%%%%%%% %%%%%%%%%%%%%%%%%%%% %%%%%%%%%%%%%%%%%%%% %%%%%%%%%%%%%%%%%%%%
 
 % \begin{widetext}
 % \begin{figure}[htbp]
 % \begin{center}
 % \epsfxsize 80mm
 % \hspace{0cm}
 % \resizebox{6in}{1.5in}{\includegraphics{MulSca.eps}} 
 % \vspace{0.25cm}
 % \caption{A heavy-quark produced (say, in DIS on a nucleon) inside a large nucleus. The produced quark then propagates through the remaining nucleus multiply scattering off the soft gluon field of the nucleons behind the struck nucleon.}
 % \label{mult-scat}
 % \end{center}
 % \end{figure}
 % \end{widetext}

 \section{Evolution of $\hat{q}$ in stochastic multiple scatterings regime}
 %%%%%%%%%%%%%%%%%%%% %%%%%%%%%%%%%%%%%%%% %%%%%%%%%%%%%%%%%%%% %%%%%%%%%%%%%%%%%%%% %%%%%%%%%%%%%%%%%%%%
 %
 Kinematic domains where in-medium interactions can be approximated by stochastic multiple scatterings, 
 ${\cal S}(r_\perp)$ appears to be a Gaussian in $r_\perp$  with variance  $2/{\hat q}L$ 
 \cite{D'Eramo:2010ak},
 \begin{eqnarray}
 {\cal S}[r_\perp] = \exp \left[- \frac{\hat q}{4} L r_\perp^2\right]~. 
 \label{priyanka}
 \end{eqnarray}
 Above form of ${\cal S}(r_\perp)$ in terms of ${\hat q}L$ and $r_\perp^2$ is limited, if not unique, 
 in the sense that its Fourier transform leaves again a Gaussian form of $P(k_\perp)$ as given in Eq. 
 \eqref{divya} and also return ${\hat q}$ once ${\hat {\cal F}}$ acts over it, 
 \begin{eqnarray}
 {\hat {\cal F}}_{[k_\perp, r_\perp]}\left[\exp \left(- \frac{\hat q}{4} L r_\perp^2\right) \right]~= {\hat q}~.
 \label{monica}
 \end{eqnarray}
 The imaginary part of the forward scattering amplitude ${\cal N}(r_\perp)$ in Glauber-Gribov-Mueller 
 (GGM) multi-rescattering model \cite{Mueller:1989st} also have a similar form, 
 \begin{eqnarray}
 {\cal N}(r_\perp) = 1-\exp\left\{-\frac{\alpha_s \pi^2}{2N_c}T(b_\perp) xG_{N}\left(x,\frac{1}{r_\perp^2}\right)~r_\perp^2\right\}~,\nn \\
 \label{lisa}
 \end{eqnarray}
 where imaginary part of the forward scattering amplitude ${\cal N}$ is related with $S$-matrix element, ${\cal S}(r_\perp)$ as, 
 \begin{eqnarray}
 {\cal N}(r_\perp)=1-{\cal S}(r_\perp)~. 
 \end{eqnarray}
 As long as one is not deep inside the saturation region Eq. \eqref{lisa} provides the correct form of 
 saturation scale  with the identification \cite{Kovchegov:2012mbw}, 
 \begin{eqnarray}
 Q_s^2(b_\perp)=\frac{\alpha_s \pi^2}{2N_c}T(b_\perp) xG_{N}\left(x,\frac{1}{r_\perp^2}\right) \sim \frac{1}{4}~{\hat q} L
 \end{eqnarray}
 In order to calculate ${\hat q}$ at strong coupling via 
 AdS/CFT correspondence, 
 form of ${\cal S}$ as given in Eq. \eqref{priyanka} was even taken to define 
 nonperturbative, quantum field theoretic definition of jet quenching parameter: 
 ${\hat q}$ is the coefficient of $Lr_\perp^2/4$ in $\log {\cal S}$ for small $r_\perp$
 \cite{Liu:2006ug}. 
 In this paper we have assumed form of all ${\cal S}$ in Eq. \eqref{kajal} as  
 Gaussian in dipoles' transverse size, with a variance of $2/{\hat q}L$, retains while it goes through high-energy evolution,
 with the caveat that deep inside the saturation region this may not true. 
 We therefore take $S$-matrices in Eq. \eqref{kajal} as, 
 \begin{eqnarray}
 {\cal S}(r_\perp, Y)&=&\exp\left[-\frac{\hat q(Y)}{4}Lr_\perp^2\right]  \label{kajal1} \\
 {\cal S}(r_\perp-B_\perp, Y)&=&\exp\left[-\frac{\hat q(Y)}{4}L(r_\perp-B_\perp)^2\right] \label{kajal2} \\
 {\cal S}(B_\perp, Y)&=&\exp\left[-\frac{\hat q(Y)}{4}LB_\perp^2\right] \label{kajal3}
 \end{eqnarray}

  Its important to recall that in this region of interest just inside the saturation region (and multiple 
 stochastic scatterings works as well) $S$-matrices are small, ${\cal S}\ll 1$ and transverse width of daughter dipoles are large  
 $B_\perp,B_\perp-r_\perp \geq 1/Q_s$. Once we replace Eq. \eqref{kajal1}-\eqref{kajal3} in Eq. \eqref{kajal}, 
 ${\cal M}(r_\perp)$ becomes, 
 \begin{eqnarray}
 {\cal M}(r_\perp) &=& \exp\left(-\frac{{\hat q}(Y)}{4}Lr_\perp^2\right) \int d^2B_\perp  
                       \frac{r_\perp^2}{(r_\perp-B_\perp)^2B_\perp^2} \nn \\
  &&           \left[1-e^{-\frac{1}{2}{\hat q}(Y)L\left(B_\perp-r_\perp\right)B_\perp}\right].     
 \label{pariniti17}
 \end{eqnarray}

  However, the Gaussian assumptions of Eq. \eqref{kajal1} - Eq. \eqref{kajal3} may not correct at low-{\it x} 
  both at small and large $r_\perp$. Small-{\it x} evolution leads to anomalous dimension modifying the 
  power of $r_\perp$ from $r_\perp^2$.  
  This means that ${\hat q}$ in Eq. \eqref{kajal1} - Eq. \eqref{kajal3} is not a function of $Y$, but more appropriately is the
  initial condition of ${\hat q}(Y)$ at $Y=0$. 
  Therefore, Eq. \eqref{pariniti17} should be understood as one step of BK evolution, where daughter dipoles 
  interact through Glauber gluons are approximated by Gaussians in Eq. \eqref{kajal1} - Eq. \eqref{kajal3}. 
 The subsequent evolution equations  should then 
  be understood as a fairly crude approximation of the actual evolution.
 %

 %We will now evaluate Eq. \eqref{pariniti17} just inside the saturation regions. 
 %
 Now, around the saturation line, the dipole size $r_\perp \sim 1/Q_{s0}$,
 and $B_\perp,(B_\perp-r_\perp) \gtrsim  1/Q_s(Y)$ (however they are still well below  $ 1/\Lambda_{\it QCD}$).
 We will further assume that transverse size of the daughter dipole ($\sim$ inverse of the gluon's transverse momentum)
 is much larger than the transverse size of the parent dipole $i.e.$ $ B_\perp,(B_\perp-r_\perp) \gg  r_\perp$. 
 With appropriate limits in the integral, Eq. \eqref{pariniti17} can then be written as,  
  \begin{eqnarray}
  {\cal M}(r_\perp)
   &=& \exp \left(- \frac{{\hat q}(Y)}{4}Lr_\perp^2\right)~\pi~r_\perp^2~ \nn \\
  &&\int_{1/Q_s^2(Y)}^{1/\Lambda_{QCD}^2}
   ~  \frac{dB_\perp^2 }{B_\perp^4}\left[1-e^{-\frac{1}{2}{\hat q}(Y)LB_\perp^2}\right]  
  \label{pariniti1}
  \end{eqnarray} 
  Integral in Eq. \eqref{pariniti1} is so converging that the upper limit may be taken at infinity instead of $1/\Lambda_{QCD}^2$,
  leaving the integral $I_{dip}$ as,      
  \begin{eqnarray}     
  I_{dip} &\equiv& \int_{1/Q_s^2(Y)}^{\infty} \frac{dB_\perp^2 }{B_\perp^4} 
     \left[1-e^{-\frac{1}{2}{\hat q}(Y)LB_\perp^2}\right]\nn \\
  &=& Q_s^2(Y)\left\{1-
   \exp\left(-\frac{1}{2}\frac{\hat{q}(Y)L}{Q_s^2(Y)}\right) \right. \nn \\
  &~&~~~~~~~~~~~~~\left. -\frac{1}{2}\frac{\hat{q}(Y)L}{Q_s^2(Y)}{\rm Er}\left(-\frac{1}{2}\frac{\hat{q}L}{Q_s^2(Y)}\right)  \right\},
   \label{pariniti5}
  \end{eqnarray}       
  where $`$Er' is the Exponential Integral function. Using Eq. \eqref{pariniti5} one can now derive the evolution equation of the jet 
  quenching parameter \label{waheeda} in stochastic multiple scatterings regime just around saturation line as,   
  \begin{eqnarray}     
  \frac{\partial {\hat q}(Y)}{\partial Y}
 &=& \frac{\alpha_s N_c}{\pi }~\frac{2Q_s^2(Y)}{L}~\left[1-
   \exp\left(-\frac{\hat{q}(Y)L}{2Q_s^2(Y)}\right) \right.   \nn \\
 &~& ~~~~~~~~~~~~~~~~~  \left.   -\frac{{\hat q}(Y)L}{2Q_s^2(Y)}~{\rm Ei}\left(-\frac{\hat{q}(Y)L}{2Q_s^2(Y)}\right)\right].\nn \\
 \label{bipasa}
 \end{eqnarray}
 Eq. \eqref{bipasa} can be written in terms of incomplete gamma fuction $\Gamma(n,a)$, 
  \begin{eqnarray}     
 && \frac{\partial \ln{\hat q}(Y)}{\partial Y}  
  = {\bar \alpha}_s~\left[\frac{1}{2\cal E}\left(1-e^{-2{\cal E}}\right)
   +\Gamma(0,2{\cal E})\right] \nn \\
 &=& {\bar \alpha}_s~\left[\ln 2 +\gamma_{E} - \ln \frac{1}{{\cal E}} + \frac{1}{2\cal E}\left(1-e^{-2{\cal E}}\right)
    + \sum_{k=1}^{\infty} \frac{\left(-{\cal E}\right)^{k}}{k (k!)} \right] \nn \\
 \label{Ananya}
 \end{eqnarray}
 where, 
 \begin{eqnarray}
  {\cal E} = \frac{1}{4} \frac{\hat{q}(Y)L}{Q_s^2(Y)}~, 
 \end{eqnarray}
 with ${\bar \alpha}_s=\alpha_s N_c / \pi $. Around ${\cal E} \sim 1$, solution of Eq. \eqref{Ananya} can be 
 approximately estimated as, 
 \begin{eqnarray}
 {\hat q(x)} \propto \left(\frac{1}{x} \right)^{0.9{\bar \alpha}_s} ~. 
 \end{eqnarray}
 Jet quenching parameter $\hat q(x)$ therefore evolves with $x$, with an exponent $0.9~{\bar \alpha}_s$, which is just slightly 
 less than that of $x {\cal G} (x)$ where ${\cal G} (x)$ is the gluon distribution function.

 \subsection{Special Case: Double log enhancement}
 For smaller dipole with $r_\perp \ll 1/Q_s$ the imaginary part of forward scattering amplitude ${\cal N}$ 
 is small, and saturation and unitarity bound effects are not that very important. 
 Principle of color transparency ensures that the forward scattering amplitude $\cal N$ 
 approaches to zero ($S$-matrix, $\cal S$,  goes to one). 
 This allows us to take that ${\cal N}\ll 1$ for small dipole size even when small-$x$ 
 evolution is included. 
 For ${\cal N}\ll 1$ we can linearise  Eq. \eqref{pariniti17} as,  
 \begin{eqnarray}
 {\cal M}(r_\perp) &=& \frac{{\hat q}(Y)}{2}L 
                       \exp\left(-\frac{{\hat q}(Y)}{4}Lr_\perp^2\right)  \nn \\
                   &~&\int d^2B_\perp  \frac{r_\perp^2}{(r_\perp-B_\perp)^2B_\perp^2}\left[(B_\perp-r_\perp)B_\perp\right]    
 \label{madhu}
 \end{eqnarray}

 \noindent In the limit of $r_\perp \ll B_\perp$, Eq. \eqref{madhu} can be further simplified, 
 \begin{eqnarray}
 {\cal M}(r_\perp) &=& \pi \frac{{\hat q}(Y)}{2}Lr_\perp^2 
                       \exp\left(-\frac{{\hat q}(Y)}{4}Lr_\perp^2\right)  
                       \int \frac{dB_\perp^2}{B_\perp^4}B_\perp^2 .  \nn \\
 \label{madhumati}
 \end{eqnarray}
 In a dipole system  
 when a single real gluon emits, the gluon can be characterized by its energy, $\omega$, and transverse
 size, $B_\perp$, that it makes with the (anti)-quark legs of the parent dipole having 
 transverse size $r_\perp$.  
 When $B_\perp \gg r_\perp$ the probability of emitting a gluon with given energy $\omega$
 goes as $r_\perp^2dB_\perp^2/B_\perp^4$. 
 Despite being non logarithmic and converging, the integral become logarithmically diverging 
 once the single scattering in included which gives an additional $B_\perp^2$ in the numerator
 \cite{Liou:2013qya}.
 This is evident from Eq. \eqref{madhumati} which together with the logarithm coming from
 energy $\omega$ integral gives double logarithmic contribution to the modification in
 transverse momentum broadening and so as for the jet quenching parameter. 
 Substituting Eq. \eqref{monica} and Eq. \eqref{madhumati} in Eq. \eqref{waheeda100} gives the $Y$ evolution of jet
 quenching  parameter, 
 \begin{eqnarray}
 \frac{\partial {\hat q}(Y)}{\partial Y} = \frac{\alpha_s N_c}{\pi} 
                \int \frac{dB_\perp^2}{B_\perp^2}
    \label{tisha}
 \end{eqnarray}
 Integral Eq. \eqref{tisha} can be compare with Eq. (26) of \cite{Liou:2013qya} 
 and Eq. (4.35) of \cite{Iancu:2014kga}. 
 This double log enhancement, first derived in \cite{Liou:2013qya} and subsequent later studies 
 \cite{Blaizot:2013vha,Iancu:2014kga,Blaizot:2014bha}, need to be supplemented with appropriate
 kinematic limits of the integral both for $B_\perp$ and $\omega$. 
 %
 %
 %\subsubsection{Kinematic limits for double log enhancement}
 
 In order to fix the kinematic limits for the energy $w$
 and $B_\perp$ here we follow the arguments made in \cite{Liou:2013qya} to 
 isolate the region of double log enhancement. 
 \noindent Two condition have to be satisfied in order to achieve the double log enhancement.  
 Firstly, the inverse transverse size of the daughter dipole $1/B_\perp^2$ ($\sim$ transverse 
 momentum of the emitted gluon) need to be just below saturation momentum $Q_s$ but sufficient
 large that the multiple scattering is not important and one can linearise Eq. \eqref{kajal}
 with confidence. 
 Secondly lifetime of the fluctuations ($\sim$ formation time of the emitted gluon) $\tau \sim \omega B_\perp^2$ 
 have to be greater than nucleon size $l_0$, and less than the length of the medium $L$. 
 Inverse transverse size of the daughter dipole is however well above the saturation scale $Q_s$ 
 which ensure $r_\perp^2 \ll B_\perp^2$. 
 \begin{eqnarray}
 r_\perp^2 \ll \frac{1}{Q_s^2}  &\leq& B_\perp^2 \leq \frac{1}{{\hat q}l_0} 
            \ll \frac{1}{\Lambda_{QCD}^2}\label{chinki} \\
 l_0 &\leq& \omega B_\perp^2 \leq L  \label{pinki}
 \end{eqnarray}
 Eq. \eqref{pinki}, together with the condition that $B_\perp$ have to be sufficiently small so that
 one can linearise Eq. \eqref{kajal}, ${\hat q} L B_\perp^2 \leq 1$, implies,  
 \begin{eqnarray}
 \frac{l_0}{B_\perp^2}  \leq  \omega  \leq \frac{1}{{\hat q} B_\perp^4}~.
 \end{eqnarray}
 \noindent As $\partial/\partial Y \equiv \omega \partial/\partial \omega$ double logarithmic 
 enhancement of jet quenching parameter in high energy would be, 
 \begin{eqnarray}
 \Delta{\hat q} &=& \frac{\alpha_s N_c}{\pi}\int_{l_0/B_\perp^2}^{1/\hat{q}(\omega)B_\perp^4} \frac{d\omega}{\omega} 
                  ~{\hat q}(\omega) 
                  ~\int_{1/Q_s^2}^{1/\hat{q}(\omega)l_0} \frac{dB_\perp^2}{B_\perp^2}
                ~  \nn \\
                &=& \frac{{\bar \alpha}_s}{2}~{\hat q}({0})~\log^2\frac{Q_s^2}{{\hat q}l_0} ~, 
              \label{alia}
 \end{eqnarray}
 \noindent where $\omega$ integration have been performed before the $B_\perp$ integration. 
 For an almost constant ${\hat q}(\omega)$ we recovered in Eq. \eqref{alia} the double log
 result (in the limit $Q_s\rightarrow {\hat q}L$) first derived in \cite{Liou:2013qya} and subsequent other studies 
 \cite{Blaizot:2013vha,Iancu:2014kga,Blaizot:2014bha}.   
 This double log enhancement however diluted by  multiple scattering effects \cite{Liou:2013qya} 
 as evident from Eq. \eqref{Ananya}.

 \section{Summary and Outlook}
 The are ongoing phenomenological efforts \ to extract values for the jet transport parameter ${\hat q}$  at various 
 central heavy-ion collisions done at various energies 
 for  prevailing energy loss models. Here parameters for the medium properties  are 
 constrained by experimental data on the nuclear modification factor $R_{AA}$ \cite{Burke:2013yra}.
 Following a recent work by D'Eramo, Liu and Rajagopal \cite{D'Eramo:2010ak} we have introduced the concept of transverse 
 deflection probability of a parton, that travels through strongly interacting medium, 
 and derived high energy evolution equation
 for the jet quenching parameter in stochastic multiple scatterings regime, which is the region of interest
 in the context of jet quenching phenomenology of the heavy-ion collider experiments.
 Balitsky-Kovchegov (BK) equation, as the evolution equation of the $S$-matrix, is used to derive high 
 energy evolution equation for the jet quenching parameter. 
 We have shown that ${\hat q}(x)$ evolves with small $x$, with an exponent $\sim 0.9~{\bar \alpha}_s$, which is just slightly 
 less than that of $x {\cal G} (x)$ where ${\cal G} (x)$ is the gluon distribution function. 
 The known result of double log enhancement emerges as a special case in the limit when the single
 scattering is only contributing. 

   In this article Eq. \eqref{kate_winslet20} - Eq. \eqref{Farida} constitute a complete set of equations needed 
   to determine the energy dependence of ${\hat q}$. To date there is no complete exact analytical solution 
   of BK equation. Although there are several approximate analytical solutions, along with some numerical solutions.
   One can therefore use the approximate analytical solutions, e.g, solution outside the saturation region 
   which exhibits extended geometric scaling or solution inside the saturation region, the Levin-Tuchin solution \cite{Levin:1999mw}, 
   that exhibits complete geometric scaling \cite{Levin:2000mv} to calculate the energy evolution of jet quenching parameter. 
   As an obvious consequence of the scaling, one could end up finding energy-dependence is
   proportional to saturation momentum, $Q_s(Y)$, $i.e.$, ${\hat q} (Y) \sim Q_s^2 (Y)$. 
 The result could therefore modify at deep inside the saturation regions and also in the event when the LPM effect modifies the spectrum strongly. 
 Both the issues which may farther squeeze the evolution of jet quenching parameter, will be explore in a future effort.

 %
 %Deep inside the saturation region  $S$-matrix, ${\cal S}$, goes to zero and the Glauber-Gribov-Mueller 
 %(GGM) amplitude ${\cal N}$ approches to unity. 
 %
 %The Balitsky-Kovchegov (BK)  equation  
 %\cite{Balitsky:1995ub,Kovchegov:1999ua,Kovchegov:1999yj,Gelis:2012ri} have a stationary point at ${\cal N}=1$ 
 %or equivalently at ${\cal S}=0$. 
 %
 %Deep inside saturation region where ${\cal N}\rightarrow1$ $({\cal S}\rightarrow0)$ non-linear evolution
 %will change the amplitude which has saturated and already reached 
 %the black-disk limit. 

 \vspace*{0.3in}

 \begin{acknowledgments}
 {\it Acknowledgments~:~} 
 %I thank Abhijit Majumder for valuable discussions during the course of this work and critically reading 
 %the manuscript.
 % 
 This work is supported by the Director, Office of Energy Research, Office of High Energy and Nuclear 
 Physics, Division of Nuclear Physics, of the U.S. Department of Energy, through the JET topical 
 collaboration. 
 \end{acknowledgments}


\begin{thebibliography}{200}
 %\cite{D'Eramo:2010ak}
  \bibitem{D'Eramo:2010ak} 
  F.~D'Eramo, H.~Liu and K.~Rajagopal,
  ``Transverse Momentum Broadening and the Jet Quenching Parameter, Redux,''
  Phys.\ Rev.\ D {\bf 84}, 065015 (2011)
  [arXiv:1006.1367 [hep-ph]].
  %%CITATION = ARXIV:1006.1367;%%
  %62 citations counted in INSPIRE as of 15 Jan 2015

  %\cite{Gelis:2012ri}
  \bibitem{Gelis:2012ri} 
  F.~Gelis,
  ``Color Glass Condensate and Glasma,''
  Int.\ J.\ Mod.\ Phys.\ A {\bf 28}, 1330001 (2013)
  [arXiv:1211.3327 [hep-ph]].
  %%CITATION = ARXIV:1211.3327;%%
  %23 citations counted in INSPIRE as of 15 Jan 2015

  %\cite{Liou:2013qya}
  \bibitem{Liou:2013qya} 
  T.~Liou, A.~H.~Mueller and B.~Wu,
  ``Radiative $p_\bot$-broadening of high-energy quarks and gluons in QCD matter,''
  Nucl.\ Phys.\ A {\bf 916}, 102 (2013)
  [arXiv:1304.7677 [hep-ph]].
  %%CITATION = ARXIV:1304.7677;%%
  %18 citations counted in INSPIRE as of 15 Jan 2015
  
  %\cite{Blaizot:2013vha}
  \bibitem{Blaizot:2013vha} 
  J.~P.~Blaizot, F.~Dominguez, E.~Iancu and Y.~Mehtar-Tani,
  ``Probabilistic picture for medium-induced jet evolution,''
  JHEP {\bf 1406}, 075 (2014)
  [arXiv:1311.5823 [hep-ph]].
  %%CITATION = ARXIV:1311.5823;%%
  %20 citations counted in INSPIRE as of 20 Jan 2015

  
  %\cite{Iancu:2014kga}
  \bibitem{Iancu:2014kga} 
  E.~Iancu,
  ``The non-linear evolution of jet quenching,''
  JHEP {\bf 1410}, 95 (2014)
  [arXiv:1403.1996 [hep-ph]].
  %%CITATION = ARXIV:1403.1996;%%
  %14 citations counted in INSPIRE as of 15 Jan 2015

  %\cite{Blaizot:2014bha}
  \bibitem{Blaizot:2014bha} 
  J.~P.~Blaizot and Y.~Mehtar-Tani,
  ``Renormalization of the jet-quenching parameter,''
  Nucl.\ Phys.\ A {\bf 929}, 202 (2014)
  [arXiv:1403.2323 [hep-ph]].
  %%CITATION = ARXIV:1403.2323;%%
  %16 citations counted in INSPIRE as of 16 Jan 2015

  %\cite{Balitsky:1995ub}
  \bibitem{Balitsky:1995ub} 
  I.~Balitsky,
  ``Operator expansion for high-energy scattering,''
  Nucl.\ Phys.\ B {\bf 463}, 99 (1996)
  [hep-ph/9509348].
  %%CITATION = HEP-PH/9509348;%%
  %1071 citations counted in INSPIRE as of 20 Jan 2015

  
  
  %\cite{Kovchegov:1999yj}
  \bibitem{Kovchegov:1999yj} 
  Y.~V.~Kovchegov,
  ``Small x F(2) structure function of a nucleus including multiple pomeron exchanges,''
  Phys.\ Rev.\ D {\bf 60}, 034008 (1999)
  [hep-ph/9901281].
  %%CITATION = HEP-PH/9901281;%%
  %948 citations counted in INSPIRE as of 20 Jan 2015

  
  %\cite{Kovchegov:1999ua}
  \bibitem{Kovchegov:1999ua} 
  Y.~V.~Kovchegov,
  ``Unitarization of the BFKL pomeron on a nucleus,''
  Phys.\ Rev.\ D {\bf 61}, 074018 (2000)
  [hep-ph/9905214].
  %%CITATION = HEP-PH/9905214;%%
  %609 citations counted in INSPIRE as of 20 Jan 2015

  %\cite{Kovchegov:2012mbw}
  \bibitem{Kovchegov:2012mbw} 
  Y.~V.~Kovchegov and E.~Levin,
  ``Quantum chromodynamics at high energy,''
  %%CITATION = INSPIRE-1217905;%%
  %13 citations counted in INSPIRE as of 21 Jan 2015

    %\cite{Dumitru:2002qt}
  \bibitem{Dumitru:2002qt} 
  A.~Dumitru and J.~Jalilian-Marian,
  ``Forward quark jets from protons shattering the colored glass,''
  Phys.\ Rev.\ Lett.\  {\bf 89}, 022301 (2002)
  [hep-ph/0204028].
  %%CITATION = HEP-PH/0204028;%%
  %130 citations counted in INSPIRE as of 07 juil. 2015

  %\cite{Dokshitzer:1977sg}
\bibitem{Dokshitzer:1977sg} 
  Y.~L.~Dokshitzer,
  ``Calculation of the Structure Functions for Deep Inelastic Scattering and 
  e+ e- Annihilation by Perturbation Theory in Quantum Chromodynamics.,''
  Sov.\ Phys.\ JETP {\bf 46}, 641 (1977)
  [Zh.\ Eksp.\ Teor.\ Fiz.\  {\bf 73}, 1216 (1977)].
  %%CITATION = SPHJA,46,641;%%
  %2705 citations counted in INSPIRE as of 07 juil. 2015

  
  %\cite{Gribov:1972ri}
\bibitem{Gribov:1972ri} 
  V.~N.~Gribov and L.~N.~Lipatov,
  ``Deep inelastic e p scattering in perturbation theory,''
  Sov.\ J.\ Nucl.\ Phys.\  {\bf 15}, 438 (1972)
  [Yad.\ Fiz.\  {\bf 15}, 781 (1972)].
  %%CITATION = SJNCA,15,438;%%
  %3062 citations counted in INSPIRE as of 07 juil. 2015

  
  
  %\cite{Altarelli:1977zs}
\bibitem{Altarelli:1977zs} 
  G.~Altarelli and G.~Parisi,
  ``Asymptotic Freedom in Parton Language,''
  Nucl.\ Phys.\ B {\bf 126}, 298 (1977).
  %%CITATION = NUPHA,B126,298;%%
  %5332 citations counted in INSPIRE as of 07 Jul 2015

  
  
  
  
  
  
  
  
  
  
  
  
  %\cite{Liang:2008vz}
  \bibitem{Liang:2008vz} 
  Z.~t.~Liang, X.~N.~Wang and J.~Zhou,
  ``The Transverse-momentum-dependent Parton Distribution Function and Jet Transport in Medium,''
  Phys.\ Rev.\ D {\bf 77}, 125010 (2008)
  [arXiv:0801.0434 [hep-ph]].
  %%CITATION = ARXIV:0801.0434;%%
  %33 citations counted in INSPIRE as of 26 Jan 2015

 
  %\cite{Mueller:1989st}
  \bibitem{Mueller:1989st} 
  A.~H.~Mueller,
  ``Small x Behavior and Parton Saturation: A QCD Model,''
  Nucl.\ Phys.\ B {\bf 335}, 115 (1990).
  %%CITATION = NUPHA,B335,115;%%
  %468 citations counted in INSPIRE as of 27 Jan 2015

  %\cite{Kovner:2003zj}
\bibitem{Kovner:2003zj}
  A.~Kovner and U.~A.~Wiedemann,
  ``Gluon radiation and parton energy loss,''
  In *Hwa, R.C. (ed.) et al.: Quark gluon plasma* 192-248
  [hep-ph/0304151].
  %%CITATION = HEP-PH/0304151;%%
  %202 citations counted in INSPIRE as of 28 Jan 2015

  %\cite{Liu:2006ug}
  \bibitem{Liu:2006ug} 
  H.~Liu, K.~Rajagopal and U.~A.~Wiedemann,
  ``Calculating the jet quenching parameter from AdS/CFT,''
  Phys.\ Rev.\ Lett.\  {\bf 97}, 182301 (2006)
  [hep-ph/0605178].
  %%CITATION = HEP-PH/0605178;%%
  %315 citations counted in INSPIRE as of 28 Jan 2015
  
  
  %\cite{Bauer:2000yr}
  \bibitem{Bauer:2000yr} 
  C.~W.~Bauer, S.~Fleming, D.~Pirjol and I.~W.~Stewart,
  ``An Effective field theory for collinear and soft gluons: Heavy to light decays,''
  Phys.\ Rev.\ D {\bf 63}, 114020 (2001)
  [hep-ph/0011336].
  %%CITATION = HEP-PH/0011336;%%
  %813 citations counted in INSPIRE as of 19 Nov 2014
 
  %\cite{Bauer:2001ct}
  \bibitem{Bauer:2001ct} 
  C.~W.~Bauer and I.~W.~Stewart,
  ``Invariant operators in collinear effective theory,''
  Phys.\ Lett.\ B {\bf 516}, 134 (2001)
  [hep-ph/0107001].
  %%CITATION = HEP-PH/0107001;%%
  %385 citations counted in INSPIRE as of 19 Nov 2014

  %\cite{Bauer:2001yt}
  \bibitem{Bauer:2001yt} 
  C.~W.~Bauer, D.~Pirjol and I.~W.~Stewart,
  ``Soft collinear factorization in effective field theory,''
  Phys.\ Rev.\ D {\bf 65}, 054022 (2002)
  [hep-ph/0109045].
  %%CITATION = HEP-PH/0109045;%%
  %678 citations counted in INSPIRE as of 19 Nov 2014
  
  %\cite{Bauer:2002nz}
  \bibitem{Bauer:2002nz} 
  C.~W.~Bauer, S.~Fleming, D.~Pirjol, I.~Z.~Rothstein and I.~W.~Stewart,
  ``Hard scattering factorization from effective field theory,''
  Phys.\ Rev.\ D {\bf 66}, 014017 (2002)
  [hep-ph/0202088].
  %%CITATION = HEP-PH/0202088;%%
  %264 citations counted in INSPIRE as of 19 Nov 2014
  
 %\cite{Burke:2013yra}
 \bibitem{Burke:2013yra} 
  K.~M.~Burke {\it et al.}  [JET Collaboration],
  ``Extracting the jet transport coefficient from jet quenching in high-energy heavy-ion collisions,''
  Phys.\ Rev.\ C {\bf 90}, no. 1, 014909 (2014)
  [arXiv:1312.5003 [nucl-th]].
  %%CITATION = ARXIV:1312.5003;%%
  %44 citations counted in INSPIRE as of 22 Apr 2015


%\cite{Levin:1999mw}
\bibitem{Levin:1999mw} 
  E.~Levin and K.~Tuchin,
  ``Solution to the evolution equation for high parton density QCD,''
  Nucl.\ Phys.\ B {\bf 573}, 833 (2000)
  [hep-ph/9908317].
  %%CITATION = HEP-PH/9908317;%%
  %220 citations counted in INSPIRE as of 07 juil. 2015

  %\cite{Levin:2000mv}
\bibitem{Levin:2000mv} 
  E.~Levin and K.~Tuchin,
  ``New scaling at high-energy DIS,''
  Nucl.\ Phys.\ A {\bf 691}, 779 (2001)
  [hep-ph/0012167].
  %%CITATION = HEP-PH/0012167;%%
  %134 citations counted in INSPIRE as of 07 juil. 2015

  
  
  
  
  
  
  
  
  
  
  
  
  
\end{thebibliography}
\end{document}